\documentclass[aip,jcp,reprint,noshowkeys,superscriptaddress]{revtex4-1}
\usepackage{graphicx,dcolumn,bm,xcolor,microtype,multirow,amscd,amsmath,amssymb,amsfonts,physics,longtable,wrapfig,txfonts}
\usepackage[version=4]{mhchem}

\usepackage[utf8]{inputenc}
\usepackage[T1]{fontenc}
\usepackage{txfonts}

\usepackage[
	colorlinks=true,
    citecolor=blue,
    breaklinks=true
	]{hyperref}
\begin{document}	

\title{Dynamical Kernels for Optical Excitations}

\author{Juliette \surname{Authier}}
	\affiliation{\LCPQ}	 
\author{Pierre-Fran\c{c}ois \surname{Loos}}
	\email{loos@irsamc.ups-tlse.fr}
	\affiliation{\LCPQ}

\begin{abstract}
We discuss the physical properties and accuracy of three distinct dynamical (\ie, frequency-dependent) kernels for the computation of optical excitations within linear response theory: 
i) an \textit{a priori} built kernel inspired by the dressed time-dependent density-functional theory (TDDFT) kernel proposed by Maitra and coworkers [\href{https://doi.org/10.1063/1.1651060}{J.~Chem.~Phys.~120, 5932 (2004)}], 
ii) the dynamical kernel stemming from the Bethe-Salpeter equation (BSE) formalism derived originally by Strinati [\href{https://doi.org/10.1007/BF02725962}{Riv.~Nuovo Cimento 11, 1--86 (1988)}], and
iii) the second-order BSE kernel derived by Yang and coworkers [\href{https://doi.org/10.1063/1.4824907}{J.~Chem.~Phys.~139, 154109 (2013)}].
The principal take-home message of the present paper is that dynamical kernels can provide, thanks to their frequency-dependent nature, additional excitations that can be associated to higher-order excitations (such as the infamous double excitations), an unappreciated feature of dynamical quantities.
We also analyze, for each kernel, the appearance of spurious excitations originating from the approximate nature of the kernels, as first evidenced by Romaniello \textit{et al.}~[\href{https://doi.org/10.1063/1.3065669}{J.~Chem.~Phys.~130, 044108 (2009)}].
Using a simple two-level model, prototypical examples of valence, charge-transfer, and Rydberg excited states are considered.
\end{abstract}

\maketitle

\section{Linear response theory}
\label{sec:LR}
Linear response theory is a powerful approach that allows to directly access the optical excitations $\omega_S$ of a given electronic system (such as a molecule) and their corresponding oscillator strengths [extracted from their eigenvectors $\T{(\bX_S \bY_S)}$] via the response of the system to a weak electromagnetic field. \cite{Oddershede_1977,Casida_1995,Petersilka_1996}
From a practical point of view, these quantities are obtained by solving non-linear, frequency-dependent Casida-like equations in the space of single excitations and de-excitations \cite{Casida_1995}
\begin{equation} \label{eq:LR}
	\begin{pmatrix}
		\bR^{\sigma}(\omega_S)	&	\bC^{\sigma}(\omega_S)
		\\
		-\bC^{\sigma}(-\omega_S)^*	&	-\bR^{\sigma}(-\omega_S)^*
	\end{pmatrix}
	\cdot 
	\begin{pmatrix}
		\bX_S^{\sigma}
		\\
		\bY_S^{\sigma}	
	\end{pmatrix}	
	=
	\omega_S
	\begin{pmatrix}
		\bX_S^{\sigma}
		\\
		\bY_S^{\sigma}	
	\end{pmatrix}		
\end{equation}
where the explicit expressions of the resonant and coupling blocks, $\bR^{\sigma}(\omega)$ and $\bC^{\sigma}(\omega)$, depend on the spin manifold ($\sigma =$ $\updw$ for singlets and $\sigma =$ $\upup$ for triplets) and the level of approximation that one employs.
Neglecting the coupling block [\ie, $\bC^{\sigma}(\omega) = 0$] between the resonant and anti-resonants parts, $\bR^{\sigma}(\omega)$ and $-\bR^{\sigma}(-\omega)^*$, is known as the Tamm-Dancoff approximation (TDA).
In the absence of symmetry breaking, \cite{Dreuw_2005} the non-linear eigenvalue problem defined in Eq.~\eqref{eq:LR} has particle-hole symmetry which means that it is invariant via the transformation $\omega \to -\omega$.
Therefore, without loss of generality, we will restrict our analysis to positive frequencies. 

In the one-electron basis of (real) spatial orbitals $\lbrace \MO{p}(\br) \rbrace$, we will assume that the elements of the matrices defined in Eq.~\eqref{eq:LR} have the following generic forms: \cite{Dreuw_2005}
\begin{subequations}
\begin{gather}
	R_{ia,jb}^{\sigma}(\omega) = (\e{a} - \e{i}) \delta_{ij} \delta_{ab} + f_{ia,jb}^{\Hxc,\sigma}(\omega)
	\\
	C_{ia,jb}^{\sigma}(\omega) = f_{ia,bj}^{\Hxc,\sigma}(\omega)
\end{gather}
\end{subequations}
where $\delta_{pq}$ is the Kronecker delta, $\e{p}$ is the one-electron (or quasiparticle) energy associated with $\MO{p}(\br)$, and
\begin{equation} \label{eq:kernel}
	f_{ia,jb}^{\Hxc,\sigma}(\omega) 
	= \iint \MO{i}(\br) \MO{a}(\br) f^{\Hxc,\sigma}(\omega) \MO{j}(\br') \MO{b}(\br') d\br d\br'
\end{equation}
Here, $i$ and $j$ are occupied orbitals, $a$ and $b$ are unoccupied orbitals, and $p$, $q$, $r$, and $s$ indicate arbitrary orbitals.
In Eq.~\eqref{eq:kernel}, 
\begin{equation} \label{eq:kernel-Hxc}
	f^{\Hxc,\sigma}(\omega) = f^{\Hx,\sigma} + f^{\co,\sigma}(\omega) 
\end{equation} 
is the (spin-resolved) Hartree-exchange-correlation (Hxc) dynamical kernel.
In the case of a spin-independent kernel, we will drop the superscript $\sigma$.
As readily seen from Eq.~\eqref{eq:kernel-Hxc}, only the correlation (c) part of the kernel is frequency dependent in a wave function context. 
However, in a density-functional context, the exchange part of the kernel can be frequency dependent if exact exchange is considered. \cite{Hesselmann_2011,Hellgren_2013}
In a wave function context, the static Hartree-exchange (Hx) matrix elements read 
\begin{equation}
	f_{ia,jb}^{\Hx,\sigma} = 2\sigma \ERI{ia}{jb} - \ERI{ib}{ja}
\end{equation} 
where $\sigma = 1$ or $0$ for singlet and triplet excited states (respectively), and 
\begin{equation}
	\ERI{pq}{rs} = \iint \MO{p}(\br) \MO{q}(\br) \frac{1}{\abs{\br - \br'}} \MO{r}(\br') \MO{s}(\br') d\br d\br'
\end{equation}
are the usual two-electron integrals. \cite{Gill_1994}
The launchpad of the present study is that, thanks to its non-linear nature stemming from its frequency dependence, a dynamical kernel potentially generates more than just single excitations.
Unless otherwise stated, atomic units are used, and we assume real quantities throughout this manuscript.

\section{The concept of dynamical quantities}
\label{sec:dyn}
As a chemist, it is maybe difficult to understand the concept of dynamical properties, the motivation behind their introduction, and their actual usefulness.
Here, we will try to give a pedagogical example showing the importance of dynamical quantities and their main purposes. \cite{Romaniello_2009b,Sangalli_2011,ReiningBook}
To do so, let us consider the usual chemical scenario where one wants to get the optical excitations of a given system.
In most cases, this can be done by solving a set of linear equations of the form
\begin{equation}
	\label{eq:lin_sys}
	\bA \cdot \bc = \omega \, \bc
\end{equation}
where $\omega$ is one of the optical excitation energies of interest and $\bc$ its transition vector.
If we assume that the matrix $\bA$ is diagonalizable and of size $N \times N$, the \textit{linear} set of equations \eqref{eq:lin_sys} yields $N$ excitation energies.
However, in practice, $N$ might be (very) large (\eg, equal to the total number of single and double excitations generated from a reference Slater determinant), and it might therefore be practically useful to recast this system as two smaller coupled systems, such that 
\begin{equation}
	\label{eq:lin_sys_split}
	\begin{pmatrix}
		\bA_1	&	\T{\bb}		\\
		\bb			&	\bA_2		\\
	\end{pmatrix}
	\cdot 
	\begin{pmatrix}
		\bc_1	\\
		\bc_2	\\
	\end{pmatrix}
	= \omega
	\begin{pmatrix}
		\bc_1	\\
		\bc_2  	\\
	\end{pmatrix}
\end{equation}
where the blocks $\bA_1$ and $\bA_2$, of sizes $N_1 \times N_1$ and $N_2 \times N_2$ (with $N_1 + N_2 =  N$), can be associated with, for example, the single and double excitations of the system.
This decomposition technique is often called L\"owdin partitioning in the literature. \cite{Lowdin_1963}

Solving separately each row of the system \eqref{eq:lin_sys_split} and assuming that $\omega \bI - \bA_2$ is invertible, we get
\begin{subequations}
\begin{gather}
	\label{eq:row1}
	\bA_1 \cdot \bc_1  + \T{\bb} \cdot \bc_2 = \omega \, \bc_1
	\\
	\label{eq:row2}
	\bc_2 = (\omega \, \bI - \bA_2)^{-1} \cdot \bb \cdot \bc_1
\end{gather}
\end{subequations}
Substituting Eq.~\eqref{eq:row2} into Eq.~\eqref{eq:row1} yields the following effective \textit{non-linear}, frequency-dependent operator
\begin{equation}
	\label{eq:non_lin_sys}
	\Tilde{\bA}_1(\omega) \cdot \bc_1 = \omega \, \bc_1
\end{equation}
with 
\begin{equation}
	\Tilde{\bA}_1(\omega) = \bA_1 +  \T{\bb} \cdot (\omega \, \bI - \bA_2)^{-1} \cdot \bb 
\end{equation}
which has, by construction, exactly the same solutions as the linear system \eqref{eq:lin_sys} but a smaller dimension.
For example, an operator $\Tilde{\bA}_1(\omega)$ built in the single-excitation basis can potentially provide excitation energies for double excitations thanks to its frequency-dependent nature, the information from the double excitations being ``folded'' into $\Tilde{\bA}_1(\omega)$ via Eq.~\eqref{eq:row2}. \cite{ReiningBook}
Note that this \textit{exact} decomposition does not alter, in any case, the values of the excitation energies.

How have we been able to reduce the dimension of the problem while keeping the same number of solutions?
To do so, we have transformed a linear operator $\bA$ into a non-linear operator $\Tilde{\bA}_1(\omega)$ by making it frequency dependent.
In other words, we have sacrificed the linearity of the system in order to obtain a new, non-linear system of equations of smaller dimension [see Eq.~\eqref{eq:non_lin_sys}]. 
This procedure converting degrees of freedom into frequency or energy dependence is very general and can be applied in various contexts. \cite{Gershgorn_1968,Malrieu_1985,LiManni_2013,Nitzsche_1978b,Davidson_1981,Rawlings_1983,Staroverov_1998,Sottile_2003,Garniron_2018,QP2,Dvorak_2019a,Dvorak_2019b}
Thanks to its non-linearity, Eq.~\eqref{eq:non_lin_sys} can produce more solutions than its actual dimension. 
However, because there is no free lunch, this non-linear system is obviously harder to solve than its corresponding linear analog given by Eq.~\eqref{eq:lin_sys}.
Nonetheless, approximations can be now applied to Eq.~\eqref{eq:non_lin_sys} in order to solve it efficiently.
For example, assuming that $\bA_2$ is a diagonal matrix is of common practice (see, for example, Ref.~\onlinecite{Garniron_2018} and references therein).

Another of these approximations is the so-called \textit{static} approximation, where one sets the frequency to a particular value. 
For example, as commonly done within the Bethe-Salpeter equation (BSE) formalism of many-body perturbation theory (MBPT), \cite{Strinati_1988} $\Tilde{\bA}_1(\omega) = \Tilde{\bA}_1 \equiv \Tilde{\bA}_1(\omega = 0)$.
In such a way, the operator $\Tilde{\bA}_1$ is made linear again by removing its frequency-dependent nature.
A similar example in the context of time-dependent density-functional theory (TDDFT) \cite{Runge_1984} is provided by the ubiquitous adiabatic approximation, \cite{Tozer_2000} which neglects all memory effects by making static the exchange-correlation (xc) kernel (\ie, frequency independent). \cite{Maitra_2012,Maitra_2016,Elliott_2011}
These approximations come with a heavy price as the number of solutions provided by the system of equations \eqref{eq:non_lin_sys} has now been reduced from $N$ to $N_1$.
Coming back to our example, in the static (or adiabatic) approximation, the operator $\Tilde{\bA}_1$ built in the single-excitation basis cannot provide double excitations anymore, and the $N_1$ excitation energies are associated with single excitations.
All additional solutions associated with higher excitations have been forever lost.
In the next section, we illustrate these concepts and the various tricks that can be used to recover some of these dynamical effects starting from the static eigenproblem.

\section{Dynamical kernels}
\label{sec:kernel}

\subsection{Exact Hamiltonian}
\label{sec:exact}

Let us consider a two-level quantum system where two opposite-spin electrons occupied the lowest-energy level. \cite{Romaniello_2009b}
In other words, the lowest orbital is doubly occupied and the system has a singlet ground state.
We will label these two orbitals, $\MO{v}$ and $\MO{c}$, as valence ($v$) and conduction ($c$) orbitals with respective one-electron Hartree-Fock (HF) energies $\e{v}$ and $\e{c}$.
In a more quantum chemical language, these correspond to the HOMO and LUMO orbitals (respectively).
The ground state $\ket{0}$ has a one-electron configuration $\ket{v\bar{v}}$, while the doubly-excited state $\ket{D}$ has a configuration $\ket{c\bar{c}}$.
There is then only one single excitation possible which corresponds to the transition $v \to c$ with different spin-flip configurations.
As usual, this produces a singlet singly-excited state $\ket{S} = (\ket{v\bar{c}} + \ket{c\bar{v}})/\sqrt{2}$, and a triplet singly-excited state $\ket{T} = (\ket{v\bar{c}} - \ket{c\bar{v}})/\sqrt{2}$. \cite{SzaboBook}

For the singlet manifold, the exact Hamiltonian in the basis of these (spin-adapted) configuration state functions reads \cite{Teh_2019}
\begin{equation} \label{eq:H-exact}
	\bH^{\updw} = 
	\begin{pmatrix}
		\mel{0}{\hH}{0}		&	\mel{0}{\hH}{S}			&	\mel{0}{\hH}{D}		\\
		\mel{S}{\hH}{0}		&	\mel{S}{\hH}{S}			&	\mel{S}{\hH}{D}		\\
		\mel{D}{\hH}{0}		&	\mel{D}{\hH}{S}			&	\mel{D}{\hH}{D}		\\
	\end{pmatrix}
\end{equation}
with 
\begin{subequations}
\begin{align}
	\mel{0}{\hH}{0} & = 2\e{v} - \ERI{vv}{vv} = \EHF
	\\
	\mel{S}{\hH  - \EHF}{S} & = \Delta\e{} + \ERI{vc}{cv} - \ERI{vv}{cc}
	\\
	\begin{split}
		\mel{D}{\hH - \EHF}{D} 
		& = 2\Delta\e{} + \ERI{vv}{vv} + \ERI{cc}{cc} 
		\\
		& + 2\ERI{vc}{cv} - 4\ERI{vv}{cc}
	\end{split}
	\\
	\mel{0}{\hH}{S} & =  0
	\\
	\mel{S}{\hH}{D} & =  \sqrt{2}[\ERI{vc}{cc} - \ERI{cv}{vv}]
	\\
	\mel{0}{\hH}{D} & = \ERI{vc}{cv}	
\end{align}
\end{subequations}
and $\Delta\e{} = \e{c} - \e{v}$.
The energy of the only triplet state is simply $\mel{T}{\hH}{T} = \EHF + \Delta\e{} - \ERI{vv}{cc}$.
Exact excitation energies are calculated as differences of these total energies.
Note that these energies are exact results within the one-electron space spanned by the basis functions.

For the sake of illustration, we will use the same molecular systems throughout this study, and consider the singlet ground state of i) the \ce{H2} molecule ($R_{\ce{H-H}} = 1.4$ bohr) in the STO-3G basis, ii) the \ce{HeH+} molecule ($R_{\ce{He-H}} = 1.4632$ bohr) in the STO-3G basis, and iii) the \ce{He} atom in Pople's 6-31G basis set. \cite{SzaboBook}
The minimal basis (STO-3G) and double-zeta basis (6-31G) have been chosen to produce two-level systems.
The STO-3G basis for two-center systems (\ce{H2} and \ce{HeH+}) corresponds to one $s$-type gaussian basis function on each center, while the 6-31G basis for the helium atom corresponds to two (contracted) $s$-type gaussian functions with different exponents.

These three systems provide prototypical examples of valence, charge-transfer, and Rydberg excitations, respectively, and will be employed to quantity the performance of the various methods considered in the present study for each type of excited states. \cite{Senjean_2015,Romaniello_2009b}
In the case of \ce{H2}, the HOMO and LUMO orbitals have $\sigma_g$ and $\sigma_u$ symmetries, respectively.
The electronic configuration of the ground state is $\sigma_g^2$, and the doubly-excited state of configuration $\sigma_u^2$ has an auto-ionising resonance nature. \cite{Bottcher_1974,Barca_2018a,Marut_2020}
The singly-excited states have $\sigma_g \sigma_u$ configurations.
In He, highly-accurate calculations reveal that the lowest doubly-excited state of configuration $1s^2$ is an auto-ionising resonance state, extremely high in energy and lies in the continuum. \cite{Madden_1963,Burges_1995,Marut_2020}
However, in a minimal basis set such as STO-3G, it is of Rydberg nature as it corresponds to a transition from a relatively compact $s$-type function to a more diffuse orbital of the same symmetry.
In the heteronuclear diatomic molecule \ce{HeH+}, a Mulliken or L\"owdin population analysis associates $1.53$ electrons on the \ce{He} center and $0.47$ electrons on the \ce{H} nucleus for the ground state. \cite{SzaboBook} 
Thus, electronic excitations in \ce{HeH+} correspond to a charge transfer from the \ce{He} nucleus to the proton.
The numerical values of the various quantities defined above are gathered in Table \ref{tab:params} for each system.

\begin{table*}
	\caption{Numerical values (in eV) of the valence and conduction orbital energies, $\e{v}$ and $\e{c}$, and two-electron integrals in the orbital basis for various two-level systems.
	\label{tab:params}
	}
	\begin{ruledtabular}
		\begin{tabular}{llcccccccc}
			System		&	Method		&	$\e{v}$			&	$\e{c}$			&	$\ERI{vv}{vv}$	&	$\ERI{cc}{cc}$	&	$\ERI{vv}{cc}$	&	$\ERI{vc}{cv}$	&	$\ERI{vv}{vc}$	&	$\ERI{vc}{cc}$	\\
			\hline
			\ce{H2}		&	HF/STO-3G	&	$-15.7282$	&	$+18.2389$	&	$+18.3566$	&	$+18.9798$	&	$+18.0565$	&	$+4.9323$	&	$0$			&	$0$	\\	
			\ce{HeH+}	&	HF/STO-3G	&	$-44.4308$	&	$-4.6935$	&	$+25.6630$	&	$+20.4773$	&	$+17.9664$	&	$+3.9565$	&	$-4.7067$	&	$+1.0145$	\\
			\ce{He}		&	HF/6-31G	&	$-24.8747$	&	$+38.0921$	&	$+27.9436$	&	$+20.8538$	&	$+23.3510$	&	$+6.1952$	&	$+8.6121$	&	$+6.9540$	\\
		\end{tabular}
	\end{ruledtabular}
\end{table*}

The exact values of the singlet single and double excitations, $\omega_{1}^{\updw}$ and $\omega_{2}^{\updw}$, and the triplet single excitation, $\omega_{1}^{\upup}$,  are reported, for example, in Table \ref{tab:Maitra}.
We are going to use these as reference for the remaining of this study.

\subsection{Maitra's dynamical kernel}
\label{sec:Maitra}

The kernel proposed by Maitra and coworkers \cite{Maitra_2004,Cave_2004} in the context of dressed TDDFT (D-TDDFT) corresponds to an \textit{ad hoc} many-body theory correction to TDDFT.
More specifically, D-TDDFT adds to the static kernel a frequency-dependent part by reverse-engineering the exact Hamiltonian: one single and one double excitations, assumed to be strongly coupled, are isolated from among the spectrum and added manually to the static kernel.
The very same idea was taken further by Huix-Rotllant, Casida and coworkers, \cite{Huix-Rotllant_2011} and tested on a large set of molecules.
Here, we start instead from a HF reference.
The static problem (\ie, the frequency-independent Hamiltonian) corresponds then to the time-dependent HF (TDHF) Hamiltonian, while in the TDA, it reduces to configuration interaction with singles (CIS). \cite{Dreuw_2005}

For the two-level model, the reverse-engineering process of the exact Hamiltonian \eqref{eq:H-exact} yields
\begin{equation} \label{eq:f-Maitra}
	f_\text{M}^{\co,\updw}(\omega) = \frac{\abs*{\mel{S}{\hH}{D}}^2}{\omega - (\mel{D}{\hH}{D} - \mel{0}{\hH}{0}) }
\end{equation}
while $f_\text{M}^{\co,\upup}(\omega) = 0$.
The expression \eqref{eq:f-Maitra} can be easily obtained by folding the double excitation onto the single excitation, as explained in Sec.~\ref{sec:dyn}.
It is clear that one must know \textit{a priori} the structure of the Hamiltonian to construct such dynamical kernel, and this obviously hampers its applicability to realistic photochemical systems where it is sometimes hard to get a clear picture of the interplay between excited states. \cite{Loos_2018a,Loos_2020b,Boggio-Pasqua_2007}

\begin{figure}
	\includegraphics[width=\linewidth]{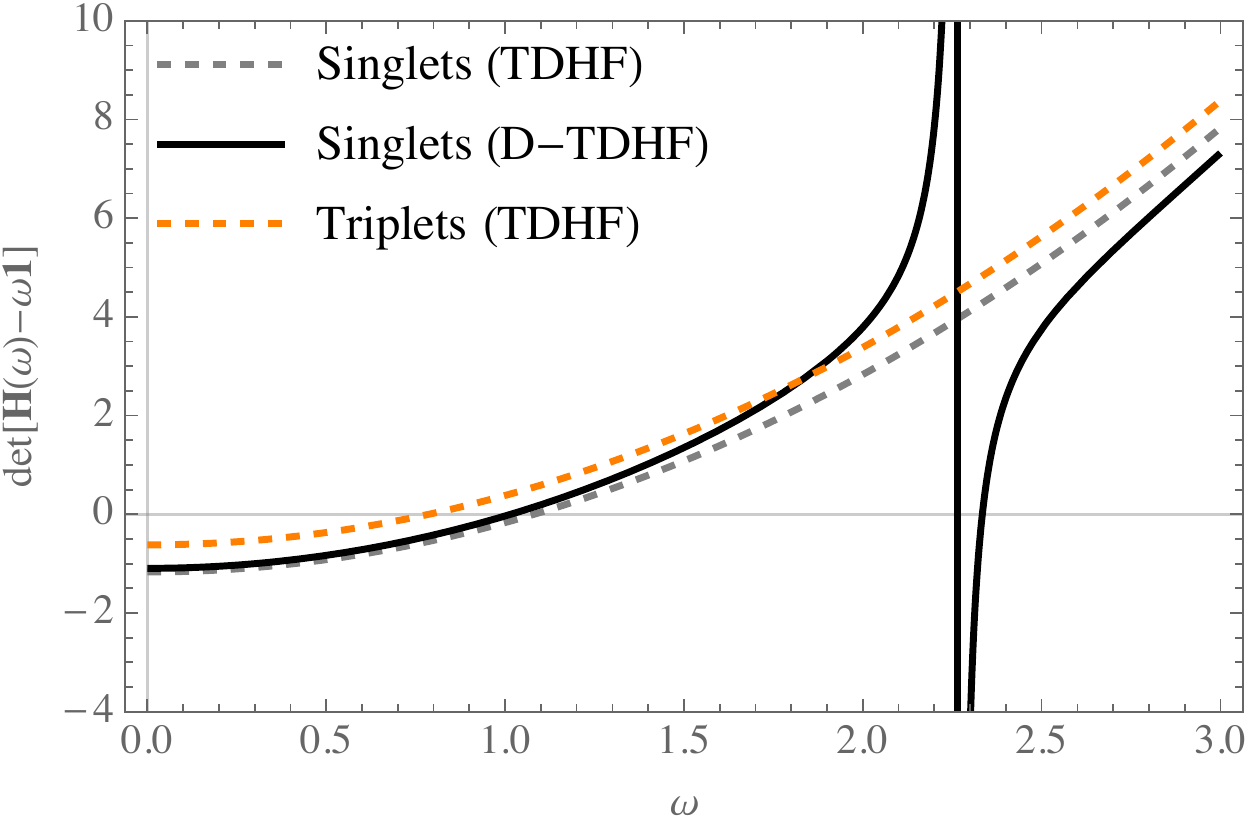}
	\caption{
	$\det[\bH(\omega) - \omega \bI]$ as a function of $\omega$ (in hartree) for both the singlet (gray and black) and triplet (orange) manifolds of \ce{HeH+}.
	The static TDHF Hamiltonian (dashed) and dynamic D-TDHF Hamiltonian (solid) are considered.
	\label{fig:Maitra}
	}
\end{figure}

For the two-level model, the non-linear equations defined in Eq.~\eqref{eq:LR} provides the following effective Hamiltonian 
\begin{equation} \label{eq:H-M}
	\bH_\text{D-TDHF}^{\sigma}(\omega) = 
	\begin{pmatrix}
		R_\text{M}^{\sigma}(\omega)		&	C_\text{M}^{\sigma}(\omega)
		\\
		-C_\text{M}^{\sigma}(-\omega)	&	-R_\text{M}^{\sigma}(-\omega)
	\end{pmatrix}	
\end{equation}
with 
\begin{subequations}
\begin{gather}
	\label{eq:R_M}
	R_\text{M}^{\sigma}(\omega) = \Delta\e{} + 2 \sigma \ERI{vc}{vc} - \ERI{vc}{vc} + f_\text{M}^{\co,\sigma}(\omega)
\\
	\label{eq:C_M}
	C_\text{M}^{\sigma}(\omega) = 2 \sigma \ERI{vc}{cv} - \ERI{vv}{cc} + f_\text{M}^{\co,\sigma}(\omega)
\end{gather}
\end{subequations}
yielding, for our three two-electron systems, the excitation energies reported in Table \ref{tab:Maitra} when diagonalized.
The TDHF Hamiltonian is obtained from Eq.~\eqref{eq:H-M} by setting $f_\text{M}^{\co,\sigma}(\omega) = 0$ in Eqs.~\eqref{eq:R_M} and \eqref{eq:C_M}.
In Fig.~\ref{fig:Maitra}, we plot $\det[\bH(\omega) - \omega \bI]$ as a function of $\omega$ for both the singlet (black and gray) and triplet (orange) manifolds in \ce{HeH+}. (Very similar curves are obtained for \ce{He}.)
The roots of $\det[\bH(\omega) - \omega \bI]$ indicate the excitation energies.
Because, there is nothing to dress for the triplet state, the TDHF and D-TDHF triplet excitation energies are equal.

\begin{table}
	\caption{Singlet and triplet excitation energies (in eV) for various levels of theory and two-level systems.
	The magnitude of the dynamical correction is reported in square brackets.
	\label{tab:Maitra}
	}
	\begin{ruledtabular}
		\begin{tabular}{lclllll}
						&						&	\mc{5}{c}{Method}	\\
													\cline{3-7}
			System		&	Excitation			&	CIS		&	TDHF	&	D-CIS			&	D-TDHF			&	Exact	\\
			\hline
			\ce{H2}		&	$\omega_1^{\updw}$	&	$25.78$	&	$25.30$	&	$25.78[+0.00]$	&	$25.30[+0.00]$	&	$26.34$	\\
						&	$\omega_2^{\updw}$	&			&			&					&					&	$44.04$	\\
						&	$\omega_1^{\upup}$	&	$15.92$	&	$15.13$	&	$15.92[+0.00]$	&	$15.13[+0.00]$	&	$16.48$	\\
			\hline
			\ce{HeH+}	&	$\omega_1^{\updw}$	&	$29.68$	&	$29.42$	&	$27.75[-1.93]$	&	$27.64[-1.78]$	&	$28.05$	\\
						&	$\omega_2^{\updw}$	&			&			&	$63.59$			&	$63.52$			&	$64.09$	\\
						&	$\omega_1^{\upup}$	&	$21.77$	&	$21.41$	&	$21.77[+0.00]$	&	$21.41[+0.00]$	&	$22.03$	\\
			\hline
			\ce{He}		&	$\omega_1^{\updw}$	&	$52.01$	&	$51.64$	&	$51.87[-0.14]$	&	$51.52[-0.12]$	&	$52.29$	\\
						&	$\omega_2^{\updw}$	&			&			&	$93.85$			&	$93.84$			&	$94.66$	\\
						&	$\omega_1^{\upup}$	&	$39.62$	&	$39.13$	&	$39.62[+0.00]$	&	$39.13[+0.00]$	&	$40.18$	\\
		\end{tabular}
	\end{ruledtabular}
\end{table}

Although not particularly accurate for the single excitations, Maitra's dynamical kernel allows to access the double excitation with good accuracy and provides exactly the right number of solutions (two singlets and one triplet).
Note that this correlation kernel is known to work best in the weak correlation regime (which is the case here) in the situation where one single and one double excitations are energetically close and well separated from the others, \cite{Maitra_2004,Loos_2019,Loos_2020d} but it is not intended to explore strongly correlated systems. \cite{Carrascal_2018}
Its accuracy for the single excitations could be certainly improved in a density-functional theory context.
However, this is not the point of the present investigation.
In Ref.~\onlinecite{Huix-Rotllant_2011}, the authors observed that the best results are obtained using a hybrid kernel for the static part.

Table \ref{tab:Maitra} also reports the slightly improved (thanks to error compensation) CIS and D-CIS excitation energies.
In particular, single excitations are greatly improved without altering the accuracy of the double excitation.
Graphically, the curves obtained for CIS and D-CIS are extremely similar to the ones of TDHF and D-TDHF depicted in Fig.~\ref{fig:Maitra}.

In the case of \ce{H2} in a minimal basis, because $\mel{S}{\hH}{D} = 0$, \cite{SzaboBook} there is no dynamical correction for both singlets and triplets, and one cannot access the double excitation with Maitra's kernel. 
It would be, of course, a different story in a larger basis set where the coupling between singles and doubles would be non-zero.
The fact that $\mel{S}{\hH}{D} = 0$ for \ce{H2} in a minimal basis is the direct consequence of the lack of orbital relaxation in the excited states, which is itself due to the fact that the molecular orbitals in that case are unambiguously defined by symmetry.

\subsection{Dynamical BSE kernel}
\label{sec:BSE}

As mentioned in Sec.~\ref{sec:dyn}, most of BSE calculations performed nowadays are done within the static approximation. \cite{ReiningBook,Onida_2002,Blase_2018,Blase_2020}
However, following Strinati's footsteps, \cite{Strinati_1982,Strinati_1984,Strinati_1988} several groups have explored this formalism beyond the static approximation by retaining the dynamical nature of the screened Coulomb potential $W$ \cite{Sottile_2003,Romaniello_2009b,Sangalli_2011,Olevano_2019} or via a perturbative approach. \cite{Rohlfing_2000,Ma_2009a,Ma_2009b,Baumeier_2012b,Loos_2020e}
Based on the very same two-level model that we employ here, Romaniello \textit{et al.} \cite{Romaniello_2009b} clearly evidenced that one can genuinely access additional excitations by solving the non-linear, frequency-dependent BSE eigenvalue problem. 
For this particular system, they showed that a BSE kernel based on the random-phase approximation (RPA) produces indeed double excitations but also unphysical excitations, \cite{Romaniello_2009b} attributed to the self-screening problem. \cite{Romaniello_2009a}
This issue was resolved in the subsequent work of Sangalli \textit{et al.} \cite{Sangalli_2011} via the design of a diagrammatic number-conserving approach based on the folding of the second-RPA Hamiltonian. \cite{Wambach_1988}
Thanks to a careful diagrammatic analysis of the dynamical kernel, they showed that their approach produces the correct number of optically active poles, and this was further illustrated by computing the polarizability of two unsaturated hydrocarbon chains (\ce{C8H2} and \ce{C4H6}).
Very recently, Loos and Blase have applied the dynamical correction to the BSE beyond the plasmon-pole approximation within a renormalized first-order perturbative treatment, \cite{Loos_2020e} generalizing the work of Rolhfing and coworkers on biological chromophores \cite{Ma_2009a,Ma_2009b} and dicyanovinyl-substituted oligothiophenes. \cite{Baumeier_2012b}
They compiled a comprehensive set of vertical transitions in prototypical molecules, providing benchmark data and showing that the dynamical correction can be sizable (especially for $n \to \pi^*$ and $\pi \to \pi^*$ excitations) and improves the static BSE excitations considerably. \cite{Loos_2020e}
Let us stress that, in all these studies, the TDA is applied to the dynamical correction (\ie, only the diagonal part of the BSE Hamiltonian is made frequency-dependent) and we shall do the same here.

Within the so-called $GW$ approximation of MBPT, \cite{Aryasetiawan_1998,Onida_2002,Reining_2017,ReiningBook,Golze_2019} one can easily compute the quasiparticle energies associated with the valence and conduction orbitals. \cite{Hybertsen_1985a,Hybertsen_1986,vanSetten_2013,Bruneval_2016}
Assuming that $W$ has been calculated at the random-phase approximation (RPA) level and within the TDA, the expression of the $\GW$ quasiparticle energy is simply \cite{Veril_2018}
\begin{equation}
	\e{p}^{\GW} = \e{p} + Z_{p}^{\GW} \SigGW{p}(\e{p})
\end{equation}
where $p = v$ or $c$, 
\begin{equation} 
	\label{eq:SigGW}
	\SigGW{p}(\omega) = \frac{2 \ERI{pv}{vc}^2}{\omega - \e{v} + \Omega} + \frac{2 \ERI{pc}{cv}^2}{\omega - \e{c} - \Omega}
\end{equation}
is the correlation part of the self-energy $\Sig{}$, and
\begin{equation}
	Z_{p}^{\GW} = \qty( 1 - \left. \pdv{\SigGW{p}(\omega)}{\omega} \right|_{\omega = \e{p}} )^{-1}
\end{equation}
is the renormalization factor (or spectral weight).
In Eq.~\eqref{eq:SigGW}, $\Omega = \Delta\e{} + 2 \ERI{vc}{cv}$ is the sole (singlet) RPA excitation energy of the system, with $\Delta\eGW{} = \eGW{c} - \eGW{v}$.

One can now build the dynamical BSE (dBSE) Hamiltonian \cite{Strinati_1988,Romaniello_2009b} 
\begin{equation} \label{eq:HBSE}
	\bH_{\dBSE}^{\sigma}(\omega) = 
	\begin{pmatrix}
		 R_{\dBSE}^{\sigma}(\omega)	&	C_{\dBSE}^{\sigma}
		\\
		-C_{\dBSE}^{\sigma}	&	-R_{\dBSE}^{\sigma}(-\omega)
	\end{pmatrix}
\end{equation}
with 
\begin{subequations}
\begin{gather}
	R_{\dBSE}^{\sigma}(\omega) = \Delta\eGW{} + 2 \sigma \ERI{vc}{cv} - \ERI{vv}{cc} - W^{\co}_R(\omega)
	\\
	C_{\dBSE}^{\sigma} = 2 \sigma \ERI{vc}{cv} -  \ERI{vc}{cv} - W^{\co}_C(\omega = 0)
\end{gather}
\end{subequations}
and where
\begin{subequations}
\begin{gather}
	\label{eq:WR}
	W^{\co}_R(\omega) = \frac{4 \ERI{vv}{vc} \ERI{vc}{cc}}{\omega - \Omega - \Delta\eGW{}}
	\\
	\label{eq:WC}
	W^{\co}_C(\omega) = \frac{4 \ERI{vc}{cv}^2}{\omega - \Omega}
\end{gather}
\end{subequations}
are the elements of the correlation part of the dynamically-screened Coulomb potential for the resonant and coupling blocks of the dBSE Hamiltonian, respectively.
Note that, in this case, the correlation kernel is spin blind.

Within the usual static approximation, the BSE Hamiltonian is simply
\begin{equation}
	\bH_{\BSE}^{\sigma} = 
	\begin{pmatrix}
		 R_{\BSE}^{\sigma} &	C_{\BSE}^{\sigma}
		\\
		-C_{\BSE}^{\sigma}	&	-R_{\BSE}^{\sigma}
	\end{pmatrix}
\end{equation}
with 
\begin{subequations}
\begin{gather}
	R_{\BSE}^{\sigma} = \Delta\eGW{} + 2 \sigma \ERI{vc}{vc} - \ERI{vv}{cc} - W_R(\omega = \Delta\eGW{})
	\\
	C_{\BSE}^{\sigma} = C_{\dBSE}^{\sigma}
\end{gather}
\end{subequations}

It can be easily shown that solving the secular equation 
\begin{equation}
	\det[\bH_{\dBSE}^{\sigma}(\omega) - \omega \bI] = 0
\end{equation}
yields 2 solutions per spin manifold (except for \ce{H2} where only one root is observed, see below), as shown in Fig.~\ref{fig:dBSE} for the case of \ce{HeH+}.
Their numerical values are reported in Table \ref{tab:BSE} alongside other variants discussed below.
These numbers evidence that dBSE reproduces qualitatively well the singlet and triplet single excitations, but quite badly the double excitation which is off by several eV.
As mentioned in Ref.~\onlinecite{Romaniello_2009b}, spurious solutions appear due to the approximate nature of the dBSE kernel. 
Indeed, diagonalizing the exact Hamiltonian \eqref{eq:H-exact} produces only two singlet solutions corresponding to the singly- and doubly-excited states, and one triplet state (see Sec.~\ref{sec:exact}). 
Therefore, there is the right number of singlet solutions but there is one spurious solution for the triplet manifold ($\omega_{2}^{\dBSE,\upup}$).
It is worth mentioning that, around $\omega = \omega_1^{\dBSE,\sigma}$, the slope of the curves depicted in Fig.~\ref{fig:dBSE} is small, while the other solution, $\omega_2^{\dBSE,\sigma}$, stems from a pole and consequently the slope is very large around this frequency value.
This makes this latter solution quite hard to locate with a method like Newton-Raphson (for example).
Let us highlight the fact that, unlike in Ref.~\onlinecite{Loos_2020e} where dynamical effects have been shown to produce a systematic red-shift of the static excitations, here we observe both blue- and red-shifted transitions (see values in square brackets in Table \ref{tab:BSE}).

\begin{table*}
	\caption{Singlet and triplet BSE excitation energies (in eV) for various levels of theory and two-level systems.
	The magnitude of the dynamical correction is reported in square brackets.
	\label{tab:BSE}
	}
	\begin{ruledtabular}
		\begin{tabular}{lclllllll}
						&						&	\mc{7}{c}{Method}	\\
													\cline{3-9}	
			System		&	Excitation			&	BSE		&	pBSE			&	dBSE			&	BSE(TDA)	&	pBSE(TDA)		&	dBSE(TDA)	&	Exact	\\
			\hline
			\ce{H2}		&	$\omega_1^{\updw}$	&	$26.06$	&	$26.06[+0.00]$	&	$26.06[+0.00]$	&	$27.02$		&	$27.02[+0.00]$	&	$27.02[+0.00]$	&	$26.34$	\\
						&	$\omega_1^{\upup}$	&	$16.94$	&	$16.94[+0.00]$	&	$16.94[+0.00]$	&	$17.16$		&	$17.16[+0.00]$	&	$17.16[+0.00]$	&	$16.48$	\\
			\hline
			\ce{HeH+}	&	$\omega_1^{\updw}$	&	$28.56$	&	$28.63[+0.07]$	&	$28.63[+0.07]$	&	$29.04$		&	$29.11[+0.07]$	&	$29.11[+0.07]$	&	$28.05$	\\
						&	$\omega_2^{\updw}$	&			&					&	$87.47$			&				&					&	$87.47$			&	$64.09$	\\
						&	$\omega_1^{\upup}$	&	$20.96$	&	$21.07[+0.11]$	&	$21.07[+0.11]$	&	$21.13$		&	$21.24[+0.11]$	&	$21.24[+0.11]$	&	$22.03$	\\
						&	$\omega_2^{\upup}$	&			&					&	$87.43$			&				&					&	$87.43$			&			\\
			\hline
			\ce{He}		&	$\omega_1^{\updw}$	&	$52.46$	&	$52.12[-0.34]$	&	$52.11[-0.35]$	&	$53.10$		&	$52.79[-0.31]$	&	$52.79[-0.31]$	&	$52.29$	\\
						&	$\omega_2^{\updw}$	&			&					&	$133.38$		&				&					&	$133.37$		&	$94.66$	\\
						&	$\omega_1^{\upup}$	&	$40.50$	&	$39.80[-0.70]$	&	$39.79[-0.71]$	&	$40.71$		&	$40.02[-0.69]$	&	$40.02[-0.69]$	&	$40.18$	\\
						&	$\omega_2^{\upup}$	&			&					&	$133.75$		&				&					&	$133.75$		&			\\
		\end{tabular}
	\end{ruledtabular}
\end{table*}

\begin{figure}
	\includegraphics[width=\linewidth]{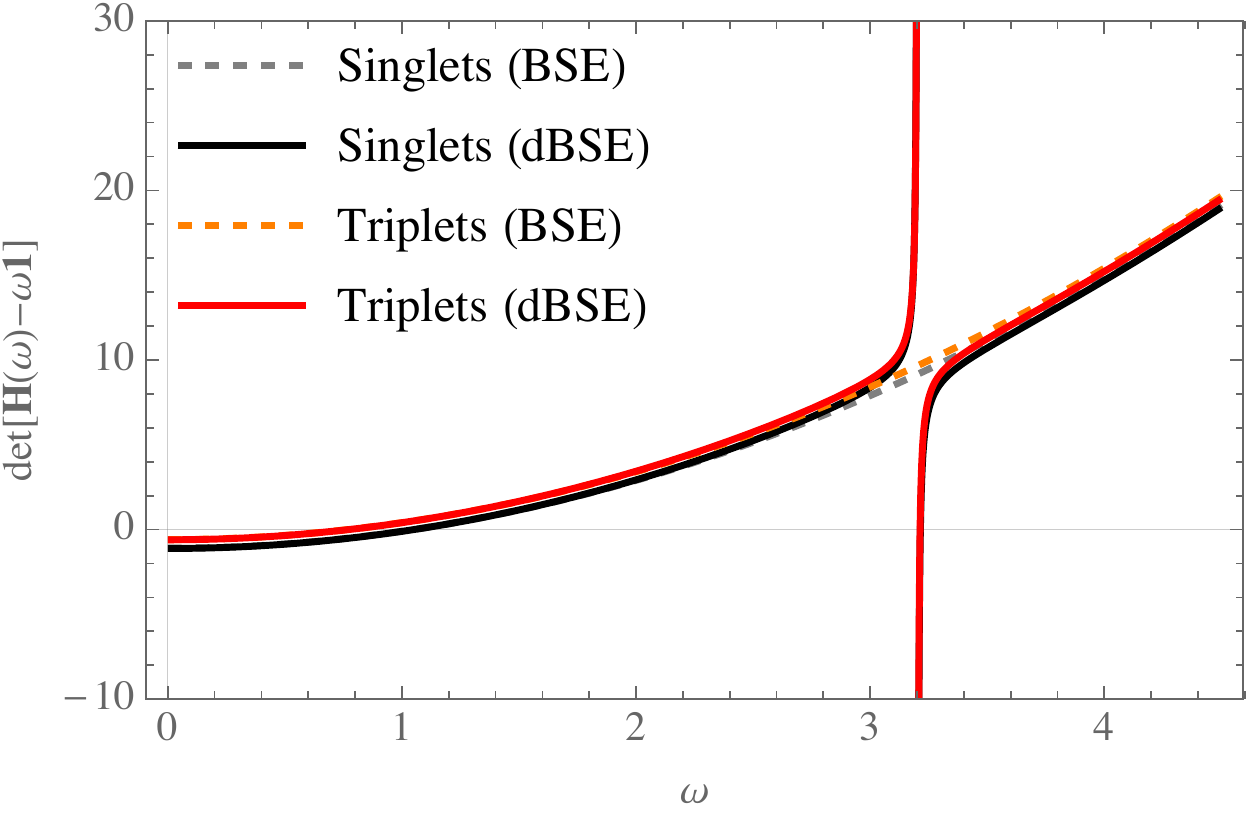}
	\caption{
	$\det[\bH(\omega) - \omega \bI]$ as a function of $\omega$ (in hartree) for both the singlet (gray and black) and triplet (orange and red) manifolds of \ce{HeH+}.
	The static BSE Hamiltonian (dashed) and dynamic dBSE Hamiltonian (solid) are considered.
	\label{fig:dBSE}
	}
\end{figure}

In the static approximation, only one solution per spin manifold is obtained by diagonalizing $\bH_{\BSE}^{\sigma}$ (see Fig.~\ref{fig:dBSE} and Table \ref{tab:BSE}).
Therefore, the static BSE Hamiltonian misses the (singlet) double excitation (as it should), and it shows that the physical single excitation stemming from the dBSE Hamiltonian is indeed the lowest in energy for each spin manifold, \ie, $\omega_1^{\dBSE,\updw}$ and $\omega_1^{\dBSE,\upup}$.
This can be further verified by switching off gradually the electron-electron interaction as one would do in the adiabatic connection formalism. \cite{Langreth_1975,Gunnarsson_1976,Zhang_2004}

Enforcing the TDA, which corresponds to neglecting the coupling term between the resonant and anti-resonant parts of the dBSE Hamiltonian \eqref{eq:HBSE}, does not change the situation in terms of spurious solutions: there is still one spurious excitation in the triplet manifold ($\omega_{2}^{\BSE,\upup}$), and the two solutions for the singlet manifold which corresponds to the single and double excitations.
However, it does increase significantly the static excitations while the magnitude of the dynamical corrections is not altered by the TDA.
The (static) BSE triplets are notably too low in energy as compared to the exact results and the TDA is able to partly reduce this error, a situation analogous in larger systems. \cite{Jacquemin_2017a,Jacquemin_2017b,Rangel_2017}

Another way to access dynamical effects while staying in the static framework is to use perturbation theory, \cite{Rohlfing_2000,Ma_2009a,Ma_2009b,Baumeier_2012b,Loos_2020e} a scheme we label as perturbative BSE (pBSE).
To do so, one must decompose the dBSE Hamiltonian into a (zeroth-order) static part and a dynamical perturbation, such that
\begin{equation}
	\bH_{\dBSE}^{\sigma}(\omega) 
	= \underbrace{\bH_{\BSE}^{\sigma}}_{\bH_{\pBSE}^{(0)}} 
	+ \underbrace{\qty[ \bH_{\dBSE}^{\sigma}(\omega) - \bH_{\BSE}^{\sigma} ]}_{\bH_{\pBSE}^{(1)}(\omega)}
\end{equation}
Thanks to (renormalized) first-order perturbation theory, \cite{Loos_2020e} one gets
\begin{equation}
\begin{split}
	\omega_{1}^{\pBSE,\sigma} 
	& = \omega_{1}^{\BSE,\sigma} 
	\\
	& + Z_{1}^{\pBSE} 
	\T{\begin{pmatrix}
		X_1	\\	Y_1
	\end{pmatrix}
	} 
	\cdot \qty[ \bH_{\dBSE}^{\sigma}(\omega =  \omega_{1}^{\BSE,\sigma}) - \bH_{\BSE}^{\sigma} ] \cdot 
	\begin{pmatrix}
		X_1	\\	Y_1
	\end{pmatrix}
\end{split}
\end{equation}
where
\begin{equation}
	\bH_{\BSE}^{\sigma}
	\cdot
	\begin{pmatrix}
		X_1	\\	Y_1
	\end{pmatrix}
	= \omega_{1}^{\BSE,\sigma}
	\begin{pmatrix}
		X_1	\\	Y_1
	\end{pmatrix}
\end{equation}
and the renormalization factor is
\begin{equation}
	Z_{1}^{\pBSE}  = \qty{ 1 - 
	\T{	
	\begin{pmatrix}
		X_1	\\	Y_1
	\end{pmatrix}
	} 
	\cdot \left. \pdv{\bH_{\dBSE}^{\sigma}(\omega)}{\omega} \right|_{\omega = \omega_{1}^{\BSE,\sigma}} \cdot 
	\begin{pmatrix}
		X_1	\\	Y_1
	\end{pmatrix}
	 }^{-1}
\end{equation}
This corresponds to a dynamical perturbative correction to the static excitations. 

The perturbatively-corrected values are also reported in Table \ref{tab:BSE}, and it shows that this scheme is very effective at reproducing the dynamical values for the single excitations.
Because the value of $Z_{1}$ is always quite close to unity in the present systems (evidencing that the perturbative expansion behaves nicely), one could have anticipated the fact that the first-order correction is a good estimate of the non-perturbative result.
However, because the perturbative treatment is ultimately static, one cannot access double excitations with such a scheme.

For \ce{H2}, there is no dynamical corrections at the BSE, pBSE or dBSE levels.
Indeed, as $\ERI{vv}{vc} = \ERI{vc}{cc} = 0$ (see Table \ref{tab:params}), we have $W^{\co}_R(\omega) = 0$ [see Eq.~\eqref{eq:WR}].
The lack of frequency dependence of the kernel means that one cannot estimate the energy of the doubly-excited state of \ce{H2}.

\subsection{Second-order BSE kernel}
\label{sec:BSE2}

The third and final dynamical kernel that we consider here is the second-order BSE (BSE2) kernel derived by Yang and collaborators in the TDA, \cite{Zhang_2013} and by Rebolini and Toulouse in a range-separated context \cite{Rebolini_2016,Rebolini_PhD} (see also Refs.~\onlinecite{Myohanen_2008,Sakkinen_2012,Olevano_2019}).
Note that a beyond-TDA BSE2 kernel was also derived in Ref.~\onlinecite{Rebolini_2016}, but was not tested.
In a nutshell, the BSE2 scheme applies second-order perturbation theory to optical excitations within the Green's function framework by taking the functional derivative of the second-order self-energy $\SigGF{}$ with respect to the one-body Green's function. 
Because $\SigGF{}$ is a proper functional derivative, it was claimed in Ref.~\onlinecite{Zhang_2013} that BSE2 does not produce spurious excitations.
However, as we will show below, this is not always true.

Like BSE requires $GW$ quasiparticle energies, BSE2 requires the second-order Green's function (GF2) quasiparticle energies, \cite{SzaboBook} which are defined as follows:
\begin{equation}
	\eGF{p} = \e{p} + Z_{p}^{\GF} \SigGF{p}(\e{p})
\end{equation}
where the second-order self-energy is
\begin{equation} 
	\label{eq:SigGF}
	\SigGF{p}(\omega) = \frac{\ERI{pv}{vc}^2}{\omega - \e{v} + \e{c} - \e{v}} + \frac{\ERI{pc}{cv}^2}{\omega - \e{c} - (\e{c} - \e{v})}
\end{equation}
and
\begin{equation}
	Z_{p}^{\GF} = \qty( 1 - \left. \pdv{\SigGF{p}(\omega)}{\omega} \right|_{\omega = \e{p}} )^{-1}
\end{equation}
The expression of the GF2 self-energy \eqref{eq:SigGF} can be easily obtained from its $GW$ counterpart \eqref{eq:SigGW} via the substitution $\Omega \to \e{c} - \e{v}$ and by dividing the numerator by a factor two. This shows that there is no screening within GF2, but that second-order exchange is properly taken into account. \cite{Zhang_2013,Loos_2018b}

The static Hamiltonian of BSE2 is just the usual TDHF Hamiltonian where one substitutes the HF orbital energies by the GF2 quasiparticle energies, \ie, 
\begin{equation}
	\bH_{\BSE2}^{\sigma} = 
	\begin{pmatrix}
		R_{\BSE2}^{\sigma}	&	C_{\BSE2}^{\sigma}
		\\
		-C_{\BSE2}^{\sigma}	&	-R_{\BSE2}^{\sigma}
	\end{pmatrix}
\end{equation}
with 
\begin{subequations}
\begin{gather}
	R_{\BSE2}^{\sigma} = \Delta\eGF{} + 2 \sigma \ERI{vc}{vc} - \ERI{vv}{cc}
	\\
	C_{\BSE2}^{\sigma} = 2 \sigma \ERI{vc}{vc} - \ERI{vc}{cv}
\end{gather}
\end{subequations}
To avoid any confusion with the results of Sec.~\ref{sec:Maitra} and for notational consistency with Sec.~\ref{sec:BSE}, we have labeled this static Hamiltonian as BSE2.

The correlation part of the dynamical kernel for BSE2 is a bit cumbersome \cite{Zhang_2013,Rebolini_2016,Rebolini_PhD} but it simplifies greatly in the case of the present model to yield
\begin{equation}
	\bH_{\dBSE2}^{\sigma} = \bH_{\BSE2}^{\sigma} +
	\begin{pmatrix}
		 R_{\dBSE2}^{\co,\sigma}(\omega)	&	C_{\dBSE2}^{\co,\sigma}
		\\
		-C_{\dBSE2}^{\co,\sigma}	&	-R_{\BSE2}^{\co,\sigma}(-\omega)
	\end{pmatrix}
\end{equation}
with 
\begin{subequations}
\begin{gather}
	R_{\dBSE2}^{\co,\updw}(\omega) = - \frac{4 \ERI{cv}{vv} \ERI{vc}{cc} - \ERI{vc}{cc}^2 - \ERI{cv}{vv}^2 }{\omega - 2 \Delta\eGF{}}
	\\
	C_{\dBSE2}^{\co,\updw} = \frac{4 \ERI{vc}{cv}^2 - \ERI{cc}{cc} \ERI{vc}{cv} - \ERI{vv}{vv} \ERI{vc}{cv} }{2 \Delta\eGF{}}
\end{gather}
\end{subequations}
and
\begin{subequations}
\begin{gather}
	R_{\dBSE2}^{\co,\upup}(\omega) = - \frac{ \ERI{vc}{cc}^2 + \ERI{cv}{vv}^2 }{\omega - 2 \Delta\eGF{}}
	\\
	C_{\dBSE2}^{\co,\upup} = \frac{\ERI{cc}{cc} \ERI{vc}{cv} + \ERI{vv}{vv} \ERI{vc}{cv} }{2 \Delta\eGF{}}
\end{gather}
\end{subequations}
As mentioned in Ref.~\onlinecite{Rebolini_2016}, the BSE2 kernel has some similarities with the second-order polarization-propagator approximation \cite{Oddershede_1977,Nielsen_1980} (SOPPA) and second RPA kernels. \cite{Wambach_1988,Huix-Rotllant_2011,Huix-Rotllant_PhD,Sangalli_2011}
Unlike the dBSE Hamiltonian [see Eq.~\eqref{eq:HBSE}], the BSE2 dynamical kernel is spin aware with distinct expressions for singlets and triplets. \cite{Rebolini_PhD} 

\begin{figure}
	\includegraphics[width=\linewidth]{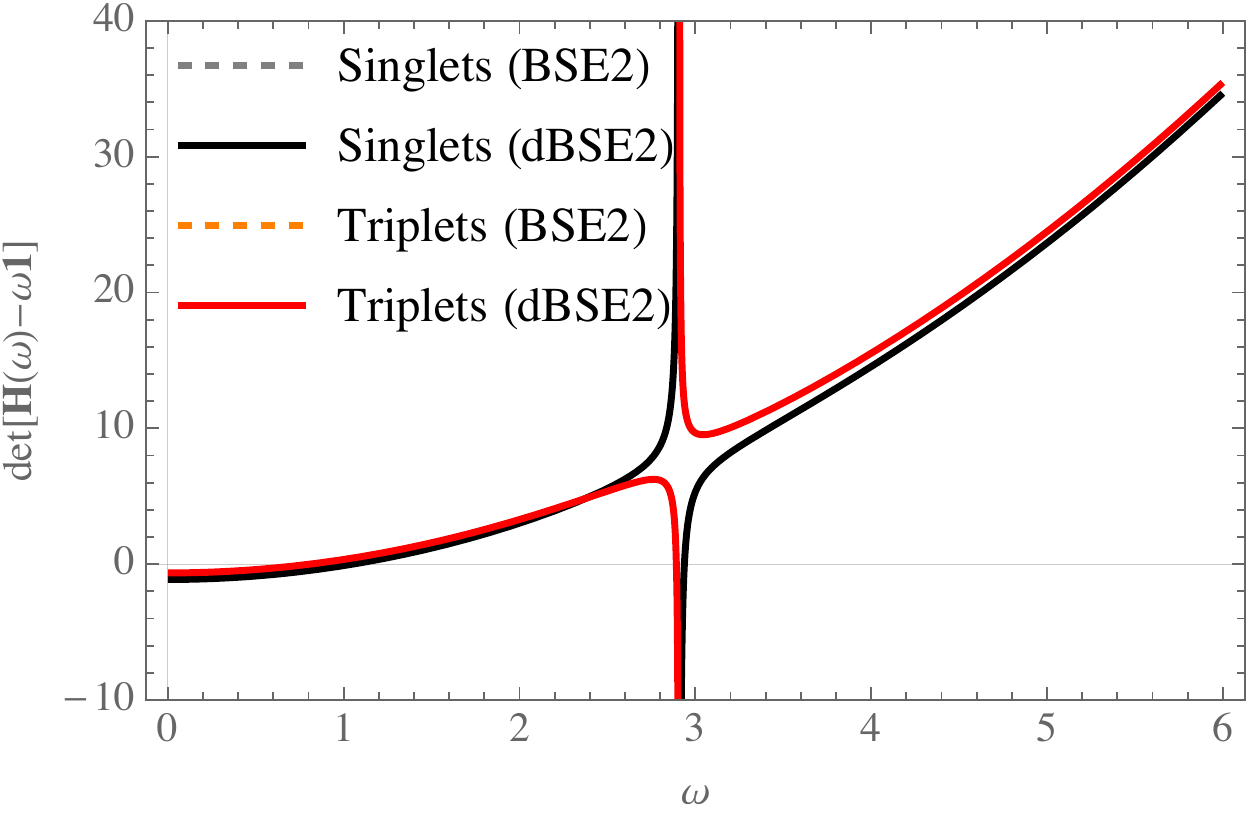}
	\caption{
	$\det[\bH(\omega) - \omega \bI]$ as a function of $\omega$ (in hartree) for both the singlet (gray and black) and triplet (orange and red) manifolds of \ce{HeH+}.
	The static BSE2 Hamiltonian (dashed) and dynamic dBSE2 Hamiltonian (solid) are considered.
	\label{fig:BSE2}
	}
\end{figure}

Like in dBSE, dBSE2 generates the right number of excitations for the singlet manifold (see Fig.~\ref{fig:BSE2}).
However, one spurious triplet excitation clearly remains.
Numerical results for the two-level models are reported in Table \ref{tab:BSE2} with the usual approximations and perturbative treatments.
In the case of BSE2, the perturbative partitioning (pBSE2) is simply
\begin{equation}
	\bH_{\dBSE2}^{\sigma}(\omega) 
	= \underbrace{\bH_{\BSE2}^{\sigma}}_{\bH_{\pBSE2}^{(0)}} 
	+ \underbrace{\qty[ \bH_{\dBSE2}^{\sigma}(\omega) - \bH_{\BSE2}^{\sigma} ]}_{\bH_{\pBSE2}^{(1)}(\omega)}
\end{equation}

\begin{table*}
	\caption{Singlet and triplet BSE2 excitation energies (in eV) for various levels of theory and two-level systems.
	The magnitude of the dynamical correction is reported in square brackets.
	\label{tab:BSE2}
	}
	\begin{ruledtabular}
		\begin{tabular}{lclllllll}
						&						&	\mc{7}{c}{Method}	\\
													\cline{3-9}
			System		&	Excitation			&	BSE2	&	pBSE2			&	dBSE2		&	BSE2(TDA)	&	pBSE2(TDA)	&	dBSE2(TDA)	&	Exact	\\
			\hline
			\ce{H2}		&	$\omega_1^{\updw}$	&	$26.03$	&	$26.03[+0.00$]	&	$26.24[+0.21]$	&	$26.49$		&	$26.49[+0.00]$	&	$26.49[+0.00]$	&	$26.34$	\\
						&	$\omega_1^{\upup}$	&	$15.88$	&	$15.88[+0.00]$	&	$16.47[+0.59]$	&	$16.63$		&	$16.63[+0.00]$	&	$16.63[+0.00]$	&	$16.48$	\\
			\hline
			\ce{HeH+}	&	$\omega_1^{\updw}$	&	$29.23$	&	$28.40[-0.83]$	&	$28.56[-0.67]$	&	$29.50$		&	$28.66[-0.84]$	&	$28.66[-0.84]$	&	$28.05$	\\
						&	$\omega_2^{\updw}$	&			&					&	$79.94$			&				&					&	$79.94$			&	$64.09$	\\
						&	$\omega_1^{\upup}$	&	$21.22$	&	$21.63[+0.41]$	&	$21.93[+0.71]$	&	$21.59$		&	$21.99[+0.40]$	&	$21.99[+0.40]$	&	$22.03$	\\
						&	$\omega_2^{\upup}$	&			&					&	$78.70$			&				&					&	$78.70$			&			\\
			\hline
			\ce{He}		&	$\omega_1^{\updw}$	&	$50.31$	&	$51.96[+1.64]$	&	$52.10[+1.79]$	&	$50.69$		&	$52.34[+1.65]$	&	$52.34[+1.65]$	&	$52.29$	\\
						&	$\omega_2^{\updw}$	&			&					&	$121.67$		&				&					&	$121.66$		&	$94.66$	\\
						&	$\omega_1^{\upup}$	&	$37.80$	&	$39.26[+1.46]$	&	$39.59[+1.79]$	&	$38.30$		&	$39.77[+1.47]$	&	$39.77[+1.47]$	&	$40.18$	\\
						&	$\omega_2^{\upup}$	&			&					&	$121.85$		&				&					&	$121.84$		&	\\
		\end{tabular}
	\end{ruledtabular}
\end{table*}

As compared to dBSE, dBSE2 produces much larger dynamical corrections to the static excitation energies, $\omega_1^{\updw}$ and $\omega_1^{\upup}$, (see values in square brackets in Table \ref{tab:BSE2}) probably due to the poorer quality of its static reference (TDHF or CIS).
Similarly to what has been observed in Sec.~\ref{sec:Maitra}, the TDA vertical excitations are slightly more accurate due to error compensations.
Note also that the perturbative treatment is a remarkably good approximation to the dynamical scheme for single excitations (except for \ce{H2}, see below), especially in the TDA. 
This justifies the use of the perturbative treatment in Refs.~\onlinecite{Zhang_2013,Rebolini_2016}.
Overall, the accuracy of dBSE and dBSE2 are comparable for single excitations although their behavior is quite different (see Tables \ref{tab:BSE} and \ref{tab:BSE2}). 
For the double excitation, dBSE2 yields a slightly better energy, yet still in quite poor agreement with the exact value.

Again, the case of \ce{H2} is a bit peculiar as the perturbative treatment (pBSE2) does not provide any dynamical corrections, while its dynamical version (dBSE2) does yield sizable corrections originating from the coupling term $C_{\dBSE2}^{\co,\sigma}$ which is non-zero in the case of dBSE2.
Although frequency-independent, this additional term makes the singlet and triplet excitation energies very accurate.
However, one cannot access the double excitation.

\section{Take-home messages}
The take-home message of the present paper is that dynamical kernels have much more to give that one would think.
In more scientific terms, dynamical kernels can provide, thanks to their frequency-dependent nature, additional excitations that can be associated to higher-order excitations (such as the infamous double excitations), an unappreciated feature of dynamical quantities.
However, they sometimes give too much, and generate spurious excitations, \ie, excitation which does not correspond to any physical excited state.
The appearance of these fictitious excitations is due to the approximate nature of the dynamical kernel.
Moreover, because of the non-linear character of the linear response problem when one employs a dynamical kernel, it is computationally more involved to access these extra excitations.

Using a simple two-model system, we have explored the physics of three dynamical kernels: i) a kernel based on the dressed TDDFT method introduced by Maitra and coworkers, \cite{Maitra_2004} ii) the dynamical kernel from the BSE formalism derived by Strinati in his hallmark 1988 paper, \cite{Strinati_1988} as well as the second-order BSE kernel derived by Zhang \textit{et al.}, \cite{Zhang_2013} and Rebolini and Toulouse. \cite{Rebolini_2016,Rebolini_PhD}
Prototypical examples of valence, charge-transfer, and Rydberg excited states have been considered.
From these, we have observed that, overall, the dynamical correction usually improves the static excitation energies, and that, although one can access double excitations, the accuracy of the BSE and BSE2 kernels for double excitations is rather average.
If one has no interest in double excitations, a perturbative treatment is an excellent alternative to a non-linear resolution of the dynamical equations.
Although it would be interesting to study the performance of such kernels in the case of stretched bonds, the appearance of singlet and triplet instabilities makes such type of investigations particularly difficult.

We hope that the present contribution will foster new developments around dynamical kernels for optical excitations, in particular to access double excitations in molecular systems.

\acknowledgements{
We would like to thank Xavier Blase, Elisa Rebolini, Pina Romaniello, Arjan Berger, Miquel Huix-Rotllant and Julien Toulouse for insightful discussions on dynamical kernels. 
PFL thanks the European Research Council (ERC) under the European Union's Horizon 2020 research and innovation programme (Grant agreement No.~863481) for financial support.

\section*{Data availability statement}
The data that supports the findings of this study are available within the article

\bibliography{dynker}

\end{document}